\documentclass[useAMS,usenatbib,usegraphicx]{mn2e}

\title[Observation of a Uranian satellite mutual event]{An observation of a mutual event between two satellites of Uranus}
\author[Hidas, Christou, \& Brown]{M.~G.~Hidas$^{1,2}$\thanks{E-mail: mhidas@lcogt.net (MGH); aac@star.arm.ac.uk (AAC)}, A.~A.~Christou$^3$\footnotemark[1], and T.~M.~Brown$^{1,2}$\\
$^{1}$Las Cumbres Observatory Global Telescope, 
6740 Cortona Dr.~Ste.~102, Goleta, CA 93117, USA\\
$^{2}$Department of Physics, University of California, 
Santa Barbara, CA 93106, USA\\
$^{3}$Armagh Observatory, 
College Hill, Armagh BT61 9DG, 
Northern Ireland, UK}
\begin{document}

\date{Accepted 2007 November 12.  Received 2007 November 9; in original form 2007 October 10}

\pagerange{\pageref{firstpage}--\pageref{lastpage}} \pubyear{2007}

\maketitle

\label{firstpage}


\begin{abstract}
We present observations of the occultation of Umbriel by Oberon on 4 May,
2007. We believe this is the first observed mutual event between satellites of
Uranus. Fitting a simple geometric model to the lightcurve, we measure the
mid-event time with a precision of 4 seconds. We assume previously measured
values for the albedos of the two satellites \citep{kar01}, and measure the
impact parameter to be $500 \pm 80$~km. These measurements are more precise
than estimates based on current ephemerides for these satellites. Therefore
observations of additional mutual events during the 2007--2008 Uranian equinox
will provide improved estimates of their orbital and physical parameters.
\end{abstract}



\section{Introduction}

The planetary satellite systems of the giant planets undergo seasons of mutual
eclipses and occultations twice during a planet's orbital revolution, when the
Sun and the Earth respectively pass through the planet's equatorial plane.
Jovian and Saturnian mutual events have been observed since 1973
\citep{aks84,arl92,arl97,thu01} resulting in very precise measurements of the
satellites' positions from so-called ``photometric astrometry''
\citep{vas03,nvd03}.

The Uranian system, although it resembles in many respects the Jovian and
Saturnian systems, has not yet benefitted from such circumstances. The last
Uranian equatorial plane crossing occurred in February 1966, well before the
advent of CCD technology.  The 2007--2008 Uranian equinox presents the only
opportunity to observe the mutual events of the Uranian satellites until the
late 2040s \citep{chr05,atl06}.  Apart from their value in improving the
satellite ephemerides and system constants, mutual event lightcurves can
provide information on large-scale albedo variations across the northern
hemispheres of the satellites that were in darkness during the Voyager 2 flyby
on Uranus in 1986 \citep{chr05}. Combined with Voyager 2 imagery, they
may enable compilation of the first global, albeit crude, albedo maps of
these bodies.
 
Here we report our observations and analysis of an occultation of Umbriel
(Uranus II) by Oberon (Uranus IV).  To our knowledge, this constitutes the
first ever observation of a mutual event between two satellites of Uranus.
 
In Section~2, we describe our observing strategy, the equipment used and the
data reduction process.  In Section~3 we present the results of our lightcurve
analysis.  We discuss the implications of our results for Uranian satellite
science in Section~4.

\section{Observations and Data reduction}

\begin{figure}
\includegraphics[width=84mm]{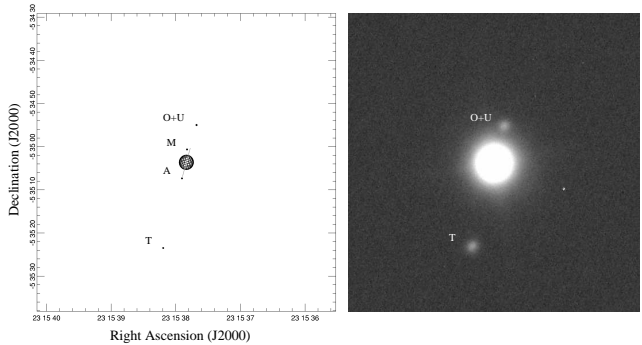}
\caption{ Predicted (left panel) and observed (right panel) configuration of
the Uranian satellites shortly after 19:00UT on 4 May, 2007.  The diagram on
the left panel was generated using M.~Showalter's {\it Uranus Viewer v2.2}
online visualisation tool ({\it pds-rings.seti.org/tools/viewer2\_ura.html}).
The image on the right shows one of the frames (Frame \#29) we acquired during
the observations, logarithmically stretched to show the moons.  Individual
satellites are indicated as follows: M -- Miranda, A -- Ariel, U -- Umbriel, T
-- Titania, O -- Oberon. At the time, the Umbriel/Oberon pair was at a
distance of 9\arcsec\ from the centre of Uranus.  Ariel and Miranda are at
4\arcsec\ and hidden in the planet's glare.  }
\label{moons}
\end{figure}

The observations were carried out on 4 May, 2007, using the Faulkes
Telescope South sited at Siding Spring, Australia and an EEV 2048x2048 CCD
with a field of view of 4.6 arcmin.  The configuration of the Uranian moons at
the time may be seen in Fig.~\ref{moons}. An SDSS~$i$ filter was used to
minimise glare from Uranus and enhance satellite contrast.  Images were binned
2x2 prior to readout resulting in an image scale of 0.27 arcsec per pixel. The
field was centred at Uranus and 3-sec exposures were acquired every $\sim$13
sec from 19:02 UT until 19:30 UT, resulting in a total of 150 frames. 

Following bias and dark subtraction and flatfielding, a fit was performed
(using pixels outside the plane occupied by the moons) to estimate the
brightness of the scattered-light halo surrounding Uranus itself.  After
subtracting this estimated stray light, differential aperture photometry was
carried out on the Umbriel-Oberon pair using Titania (Uranus III) as the
reference satellite.  The atmospheric seeing was poor and variable during the
observation, with full width at half maximum (FWHM) between 1.6 and 3.9 arcsec,
and the worst seeing occurring near the time of the occultation.  A circular
aperture with diameter of 8 pixels (2\farcs 16) gave the smallest scatter in
the relative lightcurve.

\begin{figure}
\includegraphics[width=84mm]{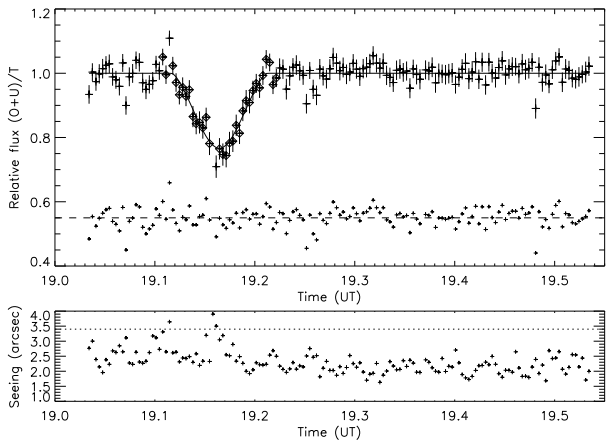}
\caption{Lightcurve of the mutual event. {\em Upper panel:} The combined flux
of Oberon and Umbriel relative to Titania as a function of Universal Time on
4 May, 2007. Diamonds indicate the points used in the fit. Also shown are the
best-fit model from CURVEFIT (solid curve) and residuals (+ symbols). {\em
Lower panel:} Estimate of the image full-width at half maximum corresponding
to each data point. The three points above the dotted line (3.4 arcsec) were
not included in the fit.}
\label{lightcurve}
\end{figure}

The three frames with the worst seeing (FWHM~$>$~3\farcs 4) resulted in
clearly discrepant relative flux values.  One of these occurs near the
beginning of the occultation, and the other two near the centre of it.
These points have been excluded from further analysis.  The lightcurve and
seeing variations can be seen in Fig.~\ref{lightcurve}.

\setbox0=\vbox{ \vspace{0cm} {DEC (DD:MM:SS)}}

Excluding the section during occultation, the relative flux time series shows a
$1\sigma$ scatter of 0.03, which is 36\% larger than the average photon shot
noise for the same interval.
The error estimates on all points were scaled up to match this larger
value. The relative fluxes were normalised to have an average value of 1
outside the event.


\section{Data Analysis and Interpretation}

A simple geometric model of the occultation was used to fit the data. The
satellites are modelled as uniformly illuminated discs with radii $R_O$
(Oberon) and $R_U$ (Umbriel). The albedo (brightness per unit disc area) of
Oberon relative to that of Umbriel is $a_{O/U}$. The combined flux of the two
satellites is then
\begin{equation}
f = 1 - \frac{A}{\pi(R_U^2 + a_{O/U}R_O^2)}
\end{equation}
where 
\begin{equation}
A = \frac{R_U^2}{2}(\theta_U - \sin\theta_U)
  + \frac{R_O^2}{2}(\theta_O - \sin\theta_O)
\end{equation}
is the area of overlap between the two discs, with
\begin{equation}
\theta_U = 2\cos^{-1}(\frac{R_U^2 + d^2 - R_O^2}{2R_Ud})
\end{equation}
(similarly for $\theta_O$, swapping the subscripts $U$ and $O$). Finally, the
distance $d$ between the centres of the two satellites (projected onto the sky)
at a time $t$ is given by
\begin{equation}
d^2(t) = x^2 + [v(t-t_0)]^2 
\end{equation}
where $x$ is the impact parameter (minimum value of $d(t)$), $v$ is the
relative speed of the two moons in the plane of the sky, and $t_0$ is the time
of maximum occultation.

The radii of the the two satellites are already known to a precision better
than 0.5\%. Thus their values in the model were fixed to $R_U = 584.7$~km and
$R_O = 761.4$~km \citep{tho88}. The relative speed was also fixed, with a value
of $v = 7.081$~km/s derived from the known orbital elements of the satellites
\citep{gio96}.

\begin{figure}
\includegraphics[width=84mm]{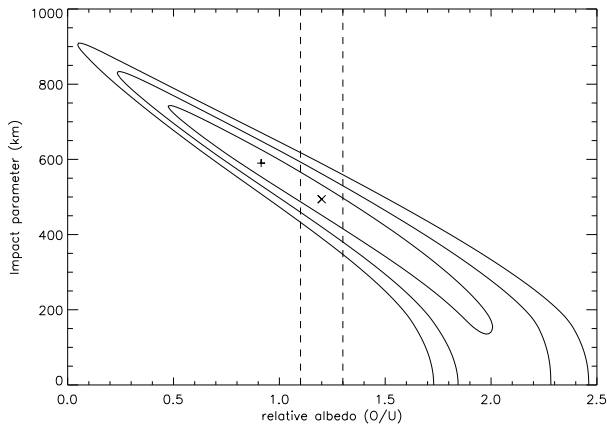}
\caption{Chi squared as a function of impact parameter and relative albedo
(Oberon / Umbriel). Contour levels correspond approximately to 1-, 2-, and
3-$\sigma$ limits. The + symbol indicates the best-fit model shown in
Fig.~\ref{lightcurve}. The vertical lines indicate the error range on an
independent estimate of the relative albedo based on data from \protect
\cite{kar01}, and X marks the best-fit impact parameter along this line.  }
\label{contour}
\end{figure}

\begin{table*}
\begin{minipage}{126mm}
  \caption{Predicted and observed parameters of the occultation of Umbriel by Oberon on 4 May, 2007. Errors in the mid-event times and measured duration are shown in brackets, in units of seconds.}
  \label{compare}  
  \begin{center}
  \begin{tabular}{lllllll}
\hline
Reference    &Ephemeris &  Event &  Mid       & Event  & Duration & Light     \\ 
             &          &  Start & Event      & End    & (sec)    & drop (\%) \\ 
\hline
\cite{chr05} & GUST86   &19:04:26&19:07:31(60)&19:10:36&  371     & 0.201 (R) \\
\cite{atl06} & LA06     &19:06:48&19:09:36(60)&19:12:24&  337     & 0.127 (R) \\
This work    &          &19:06:56&19:09:52( 4)&19:12:48&  352(10) & 0.280 (I) \\ 
\hline
  \end{tabular}
  \end{center} 
\end{minipage}
\end{table*}

The parameters to be determined from the fit are the relative albedo $a_{O/U}$,
the impact parameter $x$, and the event centre time $t_0$. The effect of the
first two on the shape of the lightcurve is symmetric about the event centre,
while the effect of changing $t_0$ is antisymmetric. As $a_{O/U}$ and $x$ both
primarily affect the depth of the signal (the latter also affects its
duration), there is a strong degeneracy between the two.

Allowing all three parameters to vary in a non-linear least-squares fit
(performed using CURVEFIT in IDL, with inverse-variance weights) gives the
values \mbox{$a_{O/U}=0.91$}, \mbox{$x=590$~km}, \mbox{$t_0=19.1645$~hours}
(19:09:52 UT). This is the fit over-plotted on the lightcurve in
Fig.~\ref{lightcurve}. The value of $t_0$ is independent of the other two
parameters and has a $1\sigma$ error of 4~seconds.

Figure~\ref{contour} shows chi squared as a function of $a_{O/U}$ and $x$ if
$t_0$ is fixed at the value above. Projecting the $1\sigma$ contour onto each
axis, the measured values with formal errors are $a_{O/U} = 0.9_{-0.4}^{+1.1}$
and $x = 600_{-450}^{+150}$~km.

An estimate for the parameter $a_{O/U}$ can be derived from independent
measurements. Table~V of \cite{kar01} lists reflectivities of the Uranian
satellites measured with the Hubble Space Telescope at various phase angles and
wavelengths. The effective wavelength for our observations was approximately
0.77~$\mu$m (SDSS~$i$ filter) and the phase angle at the time was
2.39~degrees. Averaging the tabulated values for wavelengths of 0.63~$\mu$m and
0.87~$\mu$m (at phase angle 2.82~degrees) gives reflectivities of $0.166\pm
0.007$ for Umbriel and $0.203\pm 0.009$ for Oberon, and $a_{O/U} = 1.2\pm
0.1$. From the intersection of this error range with the $1\sigma$ contour on
Fig.~\ref{contour} we obtain a more precise measurement of the impact
parameter, $x = 500 \pm 80$~km. These measurements are compared to predictions
\citep{chr05,atl06} in Table~\ref{compare}.

\section{Discussion}

We have carried out the first observation of a mutual event between two
satellites of Uranus, an occultation of Umbriel by Oberon. The parameters of
the occultation as estimated from the data have been compared to two different
sets of predictions (Table~\ref{compare}).  One employs GUST86, a Voyager-era
ephemeris while the other makes use of the more recent LA06 ephemeris which
incorporates post-1986 astrometry of the satellites. 

The errors in these predictions reflect the observational uncertainties in the
satellite positions used to derive said ephemerides. Typical
satellite-to-satellite relative positional errors of 0.03~arcseconds
\citep{chr05} translate to $\sim 400$~km at the distance of Uranus. For the
mutual event observed here, the relative velocity of the satellites is 7~km/s,
so the mid-event time predictions are uncertain by $\sim 60$~seconds. Also, the
unusual orientation of the Uranian satellite system renders precise
determination of the inclination of the orbit planes difficult when the system
is pole-on to the Earth . This was the case until the early 1990s, leading to
increased uncertainties in the predicted impact parameters.

We find that our observations are in closer agreement with the LA06
predictions. In this case, considering the above errors, the predicted and
observed mid-times are in agreement. Using \cite{kar01} to fix the relative
albedo between the two satellites, we estimate the impact parameter to be
$500\pm80$~km or $0.036\pm0.006$~arcsec compared to a value of 0.047~arcsec
predicted by LA06. The formal errors of our results are smaller than those
achieved by conventional astrometry \citep[e.g.][]{jtw98,vvm99,she02}. We thus
expect a considerable improvement in the ephemerides of the satellites to
result from observing a large number of such events predicted to occur
throughout the rest of 2007 and into 2008. This should also improve our
knowledge of some poorly-known physical parameters of the system such as the
masses of the inner three satellites Miranda, Ariel and Umbriel \citep{jac92}
and result in a better understanding of the Uranian system as a whole.

Finally, we note that our observations of this event do not strongly
constrain the relative albedo of the two satellites. This is due to the
degeneracy between the albedo and the impact parameter for a single
event. This degeneracy can be lifted either by (a) simultaneous fitting of
lightcurves for multiple events sampling different satellite aspects assuming
good a priori knowledge of the orbits or (b) a global fit of the albedo and
orbit model together.  Although both problems are sensitive to noise, they do
not contain fundamental degeneracies and have been successfully used in the
past to derive large-scale maps of Pluto \citep{you99}.  Such a fit can only
be attempted when observations of as many events as possible have been
successfully acquired.  If successful, it will yield regional to hemispherical
albedo information on the unimaged hemispheres of the major uranian
satellites.

\section*{Acknowledgements}
Astronomical research at the Armagh Observatory is funded by the Northern 
Ireland Department of Culture, Arts and Leisure (DCAL).




\bsp

\label{lastpage}

\end{document}